\documentclass{article}
\usepackage[T1]{fontenc}
\usepackage{a4wide}

\makeatletter

\newcommand{\LyX}{L\kern-.1667em\lower.25em\hbox{Y}\kern-.125emX\@}

\makeatother

\begin{document}

\textbf{Low-dimensional quantum spin systems}

\textbf{Indrani Bose}

\textbf{Department of Physics}

\textbf{Bose Institute}

\textbf{93/1, A.P.C. Road}

\textbf{Calcutta-700009}\\
(Lecture Note for the SERC School held at Mehta Research Institute,India)
\textbf{Abstract. Low-dimensional quantum spin systems are interacting many
body systems for which several rigorous results are known. Powerful techniques
like the Bethe Ansatz provide exact knowledge of the ground state energy and
low-lying excitation spectrum. A large number of compounds exists which can effectively
be described as low-dimensional spin systems. Thus many of the rigorous results
are also of experimental relevance. In this Lecture Note, an introduction is
given to quantum spin systems and the rigorous results known for such systems.
Some general theorems and exactly-solvable models are discussed. An elementary
introduction to the Bethe Ansatz technique is given.}

\section*{1.Introduction }

In Condensed Matter Physics, we study many body systems of different kinds.
In a many body system, a large number of entities interact with each other and
usually quantum effects are important. The entities may be fermions (electrons
in a metal, He\( ^{3} \) atoms ), bosons (He\( ^{4} \)atoms) or localised
spins on some lattice. In my set of lectures, we will discuss interacting spin
systems in low dimensions. The spins are located at the different sites of a
lattice. The most well-known model of interacting spins is the Heisenberg model
with the Hamiltonian

\begin{equation}
\label{1}
H=\sum _{\left\langle ij\right\rangle }J_{_{ij}}\overrightarrow{S_{i}}.\overrightarrow{S_{j}}
\end{equation}
 \( \overrightarrow{S_{i}} \) is the spin operator located at the lattice site
i . \( J_{ij} \) denotes the strength of the exchange interaction. The spin
\( \mid \overrightarrow{S_{i}} \) \( \mid  \)can have a magnitude 1/2, 1,
3/2, 2,......etc. The lattice, at the sites of which the spins are located ,
is d-dimensional. Examples are a linear chain (d = 1), the square lattice (d
= 2 ) and the cubic lattice (d = 3). Ladders have structures interpolating between
the chain and the square lattice. A n-chain ladder consists of n chains coupled
by rungs. The spin dimensionality n denotes the number of components of a spin.
The Ising, XY and Heisenberg models correspond to n =1, 2 and 3 respectively.
The generalised Ising and XY Hamiltonians are:

\begin{equation}
\label{2}
H_{Ising}=\sum _{\left\langle ij\right\rangle }J_{ij}S^{z}_{i}S^{z}_{j}
\end{equation}
 
\begin{equation}
\label{3}
H_{xy}=\sum _{\left\langle ij\right\rangle }J_{ij}\left[ S^{x}_{i}S^{x}_{j}+S^{y}_{i}S^{y}_{j}\right] 
\end{equation}
For the anisotropic XY model, \( J^{x}_{ij} \) \( \neq J_{ij}^{y} \) . Usually,
the sites i and j are nearest-neighbours (n.n.s) on the lattice and the \( J_{ij} \)
's have the same magnitude \( J \) for all the n.n. interactions. The Hamiltonian
in (1) becomes

\begin{equation}
\label{4}
H=J\sum _{\left\langle ij\right\rangle }\overrightarrow{S_{i}}.\overrightarrow{S_{j}}
\end{equation}
There are, however, other possibilities. We mention some examples :\\
 (a) Alternating chain system :

\begin{equation}
\label{5}
H=\sum _{j}\left[ 1-(-1)^{j}\delta \right] \overrightarrow{S_{j}}.\overrightarrow{S}_{j+1}
\end{equation}
 where \( 0\leq \delta \leq  \)\( 1 \).\\
(b) Further-neighbour interactions : \\
The well-known Majumdar-Ghosh chain \cite{1} is described by the Hamiltonian

\begin{equation}
\label{6}
H_{MG}=J\sum ^{N}_{i=1}\overrightarrow{S}_{i}.\overrightarrow{S}_{i+1}+J/2\sum ^{N}_{i=1}\overrightarrow{S}_{i}.\overrightarrow{S}_{i+2}
\end{equation}
 The Haldane-Shastry model\cite{2} has a Hamiltonian of the form 

\begin{equation}
\label{7}
H=J\sum _{\left\langle ij\right\rangle }\frac{1}{\mid i-j\mid ^{2}}\overrightarrow{S_{i}}.\overrightarrow{S_{j}}
\end{equation}
 (c) \( J_{ij} \) 's can be random in sign and magnitude \cite{3}.\\
(d)Quantum antiferromagnetic chain with incommensuration \cite{4}.\\
(e) Mixed-spin system: \\
A good example is a chain of alternating spins 1 and 1/2 .\\
(f) Fully anisotropic Heisenberg Hamiltonian :\\
The Hamiltonian in 1d and for n.n. interaction

\begin{equation}
\label{8}
H_{XYZ}=\sum ^{N}_{i=1}\left[ J_{x}S^{x}_{i}S^{x}_{i+1}+J_{y}S^{y}_{i}S^{y}_{i+1}+J_{z}S^{z}_{i}S^{z}_{i+1}\right] 
\end{equation}
 The Ising and XY Hamiltonians are special cases of \( H_{XYZ} \) .

Once we know the Hamiltonian of a spin system, we want to determine the eigenvalues
and the eigenfunctions of the system. This is, however, a difficult task as
the number of eigenstates is huge . For a spin-1/2 system, the number is \( 2^{N} \)
where N is the number of spins. We usually determine the ground state and the
low-lying excited states in an approximate manner. Exact knowledge can be obtained
only in a few cases. The low-lying excitation spectrum specifies the low-temperature
thermodynamics and the response to weak external fields. The spectrum may have
some universal features. In the case of electronic systems, there are two well-known
universal classes: the Fermi liquids and the Luttinger liquids. We will learn
about the universal features of spin excitation spectra as we go on. The spin
excitation spectrum can be gapless or can have a gap. In the case of a gapless
excitation spectrum, there is at least one momentum wave vector for which the
excitation energy is zero. For a spectrum with gap, the lowest excitation is
separated from the ground state by an energy gap \( \Delta  \). The temperature
dependence of thermodynamic quantities is determined by the nature of the excitation
spectrum (with or without gap). Again, one can give examples from an electronic
system. For a fermi liquid, the electronic excitation spectrum is gapless and
the electronic specific heat \( C_{n} \) has a linear temperature dependence.
A conventional superconductor is characterised by an energy gap \( \Delta  \)
in the electronic excitation spectrum and the electronic specific heat \( c_{n}\sim \exp -\frac{\Delta }{k_{B}T} \)
in the superconducting state.

The partition function Z of a thermodynamic system is

\begin{equation}
\label{9}
Z=\sum _{i}e^{-\frac{E_{i}}{k_{B}T}}
\end{equation}
where \( E_{i} \) 's are the energy levels and the sum is over all the states
of the system. The free energy

\begin{equation}
\label{10}
F=-k_{B}TlnZ
\end{equation}
 The thermodynamic quantities of a magnetic system are:\\
Magnetic energy :

\begin{equation}
\label{11}
U_{m}=\left\langle H\right\rangle =-Z^{-1}\frac{\partial Z}{\partial (1/k_{B}T)}
\end{equation}
 Entropy \( S_{H} \) at constant field H :

\begin{equation}
\label{12}
S_{H}=-\left( \frac{\partial F}{\partial T}\right) _{H}
\end{equation}
Magnetization \( M_{T} \) at constant temperature :

\begin{equation}
\label{13}
M_{T}=-\left( \frac{\partial F}{\partial H}\right) _{T}
\end{equation}
Specific heat at constant field :

\begin{equation}
\label{14}
C_{H}=\left( \frac{\partial U}{\partial T}\right) _{H}=T\left( \frac{\partial S}{\partial T}\right) _{H}=-T\left( \frac{\partial ^{2}F}{\partial T^{2}}\right) _{H}
\end{equation}
Isothermal susceptibility

\begin{equation}
\label{15}
\chi _{T}=\left( \frac{\partial M}{\partial H}\right) _{T}=-\left( \frac{\partial ^{2}F}{\partial H^{2}}\right) _{T}
\end{equation}
 Spontaneous magnetization \( M_{S} \) (H = 0) is known as the order parameter
and has a non-zero value in the ordered phase of a magnetic system. Susceptibility,
known as a response function of the system, is a measure of how the magnetization
changes as the external field H is changed. The zero-field susceptibility is

\begin{equation}
\label{16}
\chi =\left( \frac{\partial M}{\partial H}\right) _{T,H=0}
\end{equation}
At low temperatures, only the lowest energy levels are excited and a knowledge
of the low-lying excitation spectrum enables one to determine the low temperature
thermodynamics.

Exchange interaction can give rise to magnetic order below a critical temperature.
However, for some spin systems, there is no magnetic order even at T = 0. The
three major types of magnetic order are ferromagnetism (FM), antiferromagnetism
(AFM) and ferrimagnetism (Fi) . For FM order, the n.n. spins have a tendency
towards parallel alignment, for AFM order, the n.n. spins tend to be antiparallel
whereas for Fi order, the n.n. spins favour antiparallel alignment but the associated
magnetic moments have different magnitudes. In all the three cases, the magnetic
order survives upto a critical temperature \( T_{c} \) . The order parameter
in the case of an AFM is the spontaneous sublattice magnetization. Another measure
of the spontaneous order is in the two-spin correlation function. Long range
order (LRO) exists in the magnetic system if

\begin{equation}
\label{17}
lim_{R\rightarrow \infty }\left\langle \overrightarrow{S}(0).\overrightarrow{S}(\overrightarrow{R})\right\rangle \neq 0
\end{equation}
 where \( \overrightarrow{R} \) denotes the spatial location of the spin. At
T = 0, the expectation value is in the ground state and at T \( \neq  \) 0,
the expectation value is the usual thermodynamic average. 

The dynamical properties of a magnetic system are governed by the time-dependent
pair correlation functions or their Fourier transforms. An important time-dependent
correlation function is

\begin{equation}
\label{18}
G(R,t)=\left\langle \overrightarrow{S_{R}}(t).\overrightarrow{S_{0}}(0)\right\rangle 
\end{equation}
 Dynamical correlation function is the quantity measured in inelastic neutron
scattering experiments. The differential scattering cross-section in such an
experiment is given by

\begin{equation}
\label{19}
\frac{d^{2}\sigma }{d\Omega d\omega }\propto S^{\mu \mu }(\overrightarrow{q},\omega )
\end{equation}

\begin{equation}
\label{20}
=\frac{1}{N}\sum _{R}e^{i\overrightarrow{q}.\overrightarrow{R}}\int ^{+\infty }_{-\infty }dte^{i\omega t}\left\langle S^{\mu }_{R}(t)S^{\mu }_{0}(0)\right\rangle 
\end{equation}
where \( \overrightarrow{q} \) and \( \omega  \) are the momentum wave vector
and energy of the spin excitation. The spin components, \( S^{\mu } \) (\( \mu  \)
= x, y, z ), are perpendicular to \( \overrightarrow{q} \) . For a particular
\( \overrightarrow{q} \) , the peak in \( S^{\mu \mu }(\overrightarrow{q},\omega ) \)
occurs at a value of \( \omega  \) which gives the excitation energy. At T
= 0, 

\begin{equation}
\label{21}
S^{\mu \mu }(\overrightarrow{q},\omega )=\sum _{\lambda }M^{\mu }_{\lambda }\delta (\omega +E_{g}-E_{\lambda })
\end{equation}
\( E_{g} \) (\( E_{\lambda } \) ) is the energy of the ground (excited) state
and

\begin{equation}
\label{22}
M^{\mu }_{\lambda }=2\pi \mid \left\langle G\mid S^{\mu }(\overrightarrow{q})\mid \lambda \right\rangle \mid ^{2}
\end{equation}

Quantities of experimental interest include the dynamical correlation function,
various relaxation functions and associated lineshapes.

\section*{2. Ground and excited states of the FM and AFM Heisenberg Hamiltonian}

Consider the isotropic Heisenberg Hamiltonian

\begin{equation}
\label{23}
H=J\sum _{\left\langle ij\right\rangle }\overrightarrow{S_{i}}.\overrightarrow{S_{j}}
\end{equation}
where \( \left\langle ij\right\rangle  \) denotes a n.n. pair of spins. The
sign of the exchange interaction determines the favourable alignment of the
n.n. spins. J > 0 ( J < 0) corresponds to AFM (FM) exchange interaction. To
see how exchange interaction leads to magnetic order, treat the spins as classical
vectors. Each n.n. spin pair has an interaction energy \( JS^{2}\cos \theta  \)
where \( \theta  \) is the angle between n.n. spin orientations. When J is
< 0, the lowest energy is achieved when \( \theta =0 \), i.e., the interacting
spins are parallel. The ground state has all the spins parallel and the ground
state energy \( E_{g}=-\frac{JNz}{2}S^{2} \) where z is the coordination number
of the lattice. When J is > 0, the lowest energy is achieved when \( \theta =\pi  \),
i.e., the n.n. spins are antiparallel. The ground state is the Néel state in
which n.n. spins are antiparallel to each other. The ground state energy \( E_{g}=-\frac{JNz}{2}S^{2} \).

Magnetism, however, is a purely quantum phenomemon and the Hamiltonian (23)
is to be treated quantum mechanically rather than classically. For simplicity,
consider the case of a chain of spins of magnitude 1/2. Periodic boundary condition
is assumed , i.e.,\( \overrightarrow{S}_{N+1}= \) \( \overrightarrow{S}_{1} \)
.

\begin{equation}
\label{24}
H=J\sum ^{N}_{i=1}\overrightarrow{S_{i}}.\overrightarrow{S}_{i+1}=J\sum _{i=1}^{N}\left[ S^{z}_{i}S^{z}_{i+1}+\frac{1}{2}\left( S^{+}_{i}S^{-}_{i+1}+S^{-}_{i}S^{+}_{i+1}\right) \right] 
\end{equation}
where 

\begin{equation}
\label{25}
S^{\pm }_{i}=S^{x}_{i}\pm iS^{y}_{i}
\end{equation}
are the raising and lowering operators. For spins of magnitude \( \frac{1}{2} \)
, \( S^{z} \) has two possible values 1/2 and \( - \)1/2 (\( \hbar =1). \)
The corresponding states are denoted as \( \left| \alpha \right\rangle  \)
(up-spin) and \( \left| \beta \right\rangle  \) (down-spin) respectively. The
spin algebra is given by

\begin{eqnarray}
S^{z}\left| \alpha \right\rangle  & = & \frac{1}{2}\left| \alpha \right\rangle \nonumber \\
S^{z}\left| \beta \right\rangle  & = & -\frac{1}{2}\left| \beta \right\rangle \nonumber \\
S^{+}\left| \alpha \right\rangle  & = & 0\nonumber \\
S^{+}\left| \beta \right\rangle  & = & \left| \alpha \right\rangle \nonumber \\
S^{-}\left| \alpha \right\rangle  & = & \left| \beta \right\rangle \nonumber \\
S^{-}\left| \beta \right\rangle  & =0\label{26} 
\end{eqnarray}

It is easy to check that in the case of a FM, the classical ground state is
still the quantum mechanical ground state with the same ground state energy.
However, the classical AFM ground state (the Néel state) is not the quantum
mechanical ground state. The determination of the exact AFM ground state is
a tough many body problem and the solution can be obtained with the help of
the Bethe Ansatz technique (Section 4).

For a FM, the low-lying excitation spectrum can be determined exactly. Excitations
are created by deviating spins from their ground state arrangement. Deviate
a single spin in the FM ground state \( \left| \alpha \alpha \alpha \alpha ......\alpha \alpha \alpha \alpha \right\rangle  \).
The number of such states is N as the deviated spin can be at any one of the
N locations (sites). The actual eigenfunction \( \Psi  \) is a linear combination
of these states:

\begin{equation}
\label{27}
\Psi =\sum ^{N}_{m=1}a(m)\psi (m)
\end{equation}
where m denotes the location of the deviated spin and runs from 1 to N. The
wave function \( \psi (m) \) has a down spin \( \beta  \) at the mth site.

\begin{equation}
\label{28}
\psi (m)=\left| \alpha \alpha \alpha \alpha .....\beta \alpha \alpha ....\alpha \alpha \right\rangle 
\end{equation}
 To solve the eigenvalue problem

\begin{equation}
\label{29}
H\Psi =E\Psi 
\end{equation}
one has to determine the unknown coefficients a(m).

\begin{equation}
\label{30}
H\sum _{m}a(m)\psi (m)=E\sum _{m}a(m)\psi (m)
\end{equation}
Multiply by \( \psi ^{*}\left( l\right)  \) on both sides to get 

\begin{eqnarray}
\sum _{m}a(m)\left\langle \psi (l)\right| H\left| \psi (m)\right\rangle  & = & E\sum _{m}a(m)\left\langle \psi (l)\right| \left. \psi (m)\right\rangle \nonumber \\
 & = & E\sum _{m}a(m)\delta _{ml}\nonumber \\
 & = & Ea(l)\label{31} 
\end{eqnarray}

On the l.h.s., only those terms survive for which \( H\left| \psi (m)\right\rangle =\left| \psi (l)\right\rangle  \).
Consider the FM version of the Heisenberg Hamiltonian in (24),i.e., replace
J by -J with J> 0. The Hamiltonian H is a sum of two parts:

\begin{equation}
\label{32}
H=H_{z}+H_{XY}
\end{equation}
We consider the effect of each Hamiltonian on a n.n. spin pair. The following
rules hold true. \( H_{z} \) acting on a parallel (antiparallel) spin pair
gives -J/4 (J/4) times the same spin pair.

\begin{equation}
\label{33}
-JS^{z}_{i}S^{z}_{i+1}\left| \alpha \alpha \right\rangle =-\frac{J}{4}\left| \alpha \alpha \right\rangle ,-JS^{z}_{i}S^{z}_{i+1}\left| \alpha \beta \right\rangle =\frac{J}{4}\left| \alpha \beta \right\rangle 
\end{equation}
\( H_{XY} \) acting on a parallel spin pair gives zero and acting on an antiparallel
spin pair interchanges the spins with the coefficient \( - \)J/2

\begin{eqnarray}
-\frac{J}{2}(S^{+}_{i}S^{-}_{i+1}+S^{-}_{i}S^{+}_{i+1})\left| \alpha \alpha \right\rangle  & = & 0\nonumber \\
-\frac{J}{2}(S^{+}_{i}S^{-}_{i+1}+S^{-}_{i}S^{+}_{i+1})\left| \alpha \beta \right\rangle  & = & -\frac{J}{2}\left| \beta \alpha \right\rangle \nonumber \\
-\frac{J}{2}(S^{+}_{i}S^{-}_{i+1}+S^{-}_{i}S^{+}_{i+1})\left| \beta \alpha \right\rangle  & = & -\frac{J}{2}\left| \alpha \beta \right\rangle \label{34} 
\end{eqnarray}

In the wave function \( \psi  \)(m) , there are \( (N-2) \) parallel and 2
antiparallel pairs of spins. Applying the rules of spin algebra, we get from
(31),

\begin{eqnarray}
Ea(l) & = & \left[ -\frac{J}{4}(N-2)+\frac{J}{4}\times 2\right] a(l)-\frac{J}{2}\left[ a(l+1)+a(l-1)\right] \nonumber \\
\epsilon a(l) & =\frac{J}{2} & \left[ 2a(l)-a(l+1)-a(l-1)\right] \label{35} 
\end{eqnarray}
 where, \( \epsilon  \) = E + \( \frac{JN}{4} \) , is the energy of the excited
state measured w.r.t. the ground state energy \( E_{g}=-\frac{JN}{4} \) . The
solution for \( a(l) \) in (35) is

\begin{equation}
\label{36}
a(l)=e^{ikl}
\end{equation}
and we get

\begin{equation}
\label{37}
\epsilon =J(1-cos(k))
\end{equation}
 From the PBC,

\begin{equation}
\label{38}
a(l)=a(l+N)
\end{equation}
 which leads to

\begin{equation}
\label{39}
e^{ikN}=1
\end{equation}
 or, \( k=\frac{2\pi }{N}\lambda ,\lambda =0,1,2,.....,N-1 \) .\\
The excited state consists of a deviated down spin propagating along the chain.
The excitation is called a spin wave or a magnon. For r down spins (r = 2,3,4,...
etc.), we have r magnon excitations. The r magnons can scatter against each
other giving rise to a continuum of scattering states or they can form bound
states. The exact bound state spectrum can be derived using the BA (Section
4). The case of the AFM Heisenberg Hamiltonian will be considered in this Section.

The BA technique is applicable to quantum many body systems only in 1d. However,
not all 1d quantum many body problems can be solved by the BA. For such problems
as well as for models in higher dimensions, approximate techniques are used
to gain knowledge of the ground state and the excitation spectrum \cite{5,6}.\\
\textbf{3. Theorems and rigorous results for Heisenberg AFMs}\\
\textbf{A. Lieb-Mattis theorem \cite{7}}: For general spin and for all dimensions
and also for a bipartite lattice, the entire eigenvalue spectrum satisfies the
inequality

\[
E_{0}(S)\leq E_{0}(S+1)\]
 where \( E_{0}(S) \) is the minimum energy corresponding to total spin S.
The weak inequality becomes a strict inequality for a FM exchange coupling between
spins of the same sublattice. The theorem is valid for any range of exchange
coupling and the proof does not require PBC. The ground state of the S=1/2 Heisenberg
AFM with an even number N of spins is a singlet according to the Lieb-Mattis
theorem.\\
\textbf{B. Marshall's sign rule \cite{6,8}:} \\
The rule specifies the structure of the ground state of a n.n. S = 1/2 Heisenberg
Hamiltonian defined on a bipartite lattice. The rule can be generalised to spin
S, n.n.n. FM interaction but not to n.n.n. AFM interaction. A bipartite lattice
is a lattice which can be divided into two sublattices A and B such that the
n.n. spins of a spin belonging to the A sublattice are located in the B sublattice
and vice versa. Examples of such lattices are the linear chain, the square and
the cubic lattices. According to the sign rule, the ground state \( \psi  \)
has the form

\begin{equation}
\label{40}
\left| \psi \right\rangle =\sum _{\mu }C_{\mu }\left| \mu \right\rangle 
\end{equation}
where \( \left| \mu \right\rangle  \) is an Ising basis state. The coefficient
\( C_{\mu } \) has the form

\begin{equation}
\label{41}
C_{\mu }=(-)^{p_{\mu }}a_{\mu }
\end{equation}
 with \( a_{\mu } \) real and \( \geq  \)0 and \( p_{\mu } \) is the number
of up-spins in the A sublattice.\\
Proof: We write the AFM Heisenberg Hamiltonian in the form

\begin{eqnarray}
H & = & 2J\sum _{<ij>}(\overrightarrow{S}_{i}.\overrightarrow{S}_{j}-\frac{1}{4})\nonumber \\
 & = & 2J\sum _{<ij>}(S^{z}_{i}S^{z}_{j}-\frac{1}{4})+J\sum _{<ij>}\left[ S^{+}_{i}S^{-}_{j}+S^{-}_{i}S^{+}_{j}\right] \label{42} \\
 & = & H_{z}+H_{XY}\label{43} 
\end{eqnarray}
 Let \( m_{\mu } \) be the z-component of the total spin and \( q_{\mu } \)
the number of antiparallel bonds in the spin state \( \left| \mu \right\rangle  \).
\( H_{z} \) acting on a parallel spin pair gives zero and acting on an antiparallel
spin pair gives \( -J \). \( H_{XY} \) acting on a parallel spin pair gives
zero and interchanges the spins in the case of an antiparallel spin pair. In
the latter case, \( m_{\mu } \) and \( q_{\mu } \) are unchanged and \( p_{\mu }\rightarrow p_{\mu }\pm 1 \)
. 

\begin{equation}
\label{44}
H\left| \psi \right\rangle =J\sum _{\mu }C_{\mu }\{-q_{\mu }\left| \mu \right\rangle +\sum _{\nu }\left| \nu \right\rangle \}
\end{equation}
where \( \left| \nu \right\rangle  \) differs from \( \left| \mu \right\rangle  \)
by one interchange of antiparallel n.n.s. There are \( q_{\mu } \) such states.

\begin{eqnarray}
E_{c} & = & \frac{\left\langle \psi \mid H\mid \psi \right\rangle }{\left\langle \psi \mid \psi \right\rangle }\nonumber \\
 & = & \frac{J\sum _{\mu }\{-q_{\mu }C_{\mu }^{*}C_{\mu }+\sum _{\nu }C_{\nu }^{*}C_{\mu }\}}{\sum _{\mu }\mid C_{\mu }\mid ^{2}}\nonumber \\
 & = & \frac{J\sum _{\mu }\sum _{\nu }C_{\mu }\left[ -C_{\mu }^{*}+C_{\nu }^{*}\right] }{\sum _{\mu }\mid C_{\mu }\mid ^{2}}\label{45} 
\end{eqnarray}
 Let

\begin{equation}
\label{46}
C_{\mu }=(-1)^{p_{\mu }}a_{\mu }
\end{equation}
where \( a_{\mu } \) is arbitrary. Since \( p_{\nu }=p_{\mu }\pm 1 \), from
(46),

\begin{eqnarray}
C_{\mu }C^{*}_{\nu } & = & (-1)^{2p_{\mu }\pm 1}a_{\mu }a^{*}_{\nu }\nonumber \\
 & = & -a_{\mu }a_{\nu }^{*}\label{47} 
\end{eqnarray}
Also, \( C_{\mu }C_{\mu }^{*}=a_{\mu }a_{\mu }^{*} \). From (45),

\begin{equation}
\label{48}
E_{c}=-J\frac{\sum _{\mu }\sum _{\nu }a_{\mu }\left[ a^{*}_{\mu }+a_{\nu }^{*}\right] }{\sum _{\mu }a_{\mu }^{*}a_{\mu }}
\end{equation}
Now, since \( \mu ,\nu  \) are dummy indices

\begin{equation}
\label{49}
\sum _{\mu }\sum _{\nu }a_{\mu }a_{\nu }^{*}=\frac{1}{2}\sum _{\mu }\sum _{\nu }\left[ a_{\mu }a_{\nu }^{*}+a_{\mu }^{*}a_{\nu }\right] 
\end{equation}
If \( a_{\mu ,}a_{\nu } \) are real and positive, \( a_{\mu }\rightarrow \mid a_{\mu }\mid  \),
\( a_{\nu }\rightarrow \mid a_{\nu }\mid  \)and

\begin{equation}
\label{50}
a_{\mu }a_{\nu }=\mid a_{\mu }\mid \mid a_{\nu }\mid 
\end{equation}
If \( a_{\mu ,}a_{\nu } \) are complex, the term in (49) is 

\( \frac{1}{2}\{\mid a_{\mu }\mid e^{i\phi _{\mu }}\mid a_{\nu }\mid e^{-i\phi _{\nu }}+\mid a_{\mu }\mid e^{-i\phi _{\mu }}\mid a_{\nu }\mid e^{i\phi _{\nu }}\}=\mid a_{\mu }\mid \mid a_{\nu }\mid cos(\phi _{\mu }-\phi _{\nu }) \)

\[
\leq \mid a_{\mu }\mid \mid a_{\nu }\mid \]
The energy \( E_{c} \) in (45) can be reduced by choosing \( a_{\mu } \)'s
to be real and \( \geq  \)0 (proved).\\
\textbf{C. Lieb, Schultz and Mattis theorem\cite{6,9}}:\\
A half-integer S spin chain described by a reasonably local Hamiltonian respecting
translational and rotational symmetry either has gapless excitation spectrum
or has degenerate ground states, corresponding to spontaneously broken translational
symmetry. \\
Proof: We consider the Heisenberg AFM Hamiltonian in (24). A rigorous proof
of uniqueness (non-degeneracy) of the ground state \( \left| \psi _{0}\right\rangle  \)
exists in this case. We wish to prove that there is a low energy excitation
of O(1/L), where L is the length of the chain. The proof consists of two steps
:\\
(i) Construct a state \( \left| \psi _{1}\right\rangle  \) which has low energy,
i.e.,

\begin{equation}
\label{51}
\left\langle \psi _{1}\right| (H-E_{0})\left| \psi _{0}\right\rangle =O(1/L)
\end{equation}
(ii) Show that \( \left| \psi _{1}\right\rangle  \) is orthogonal to the ground
state \( \left| \psi _{0}\right\rangle  \). The state \( \left| \psi _{1}\right\rangle  \)
is constructed by applying an unitary transformation on \( \left| \psi _{0}\right\rangle  \)
which gives rise to a slowly varying rotation of the spins about the z-axis. 

\begin{eqnarray}
\left| \psi _{1}\right\rangle  & = & U\left| \psi _{0}\right\rangle \nonumber \\
U & = & exp\left[ i\frac{2\pi }{L}\sum _{n}nS_{n}^{z}\right] \label{52} 
\end{eqnarray}
The operator U rotates the nth spin by the small angle \( \frac{2\pi n}{L} \)about
the z-axis. The spin coordinates are rotated by \( 2\pi  \) about the z-axis
between the first and last sites. Note the spin identities

\begin{eqnarray}
e^{i\theta S^{z}}S^{+}e^{-i\theta S^{z}} & = & e^{i\theta }S^{+}\nonumber \\
e^{i\theta S^{z}}S^{-}e^{-i\theta S^{z}} & = & e^{-i\theta }S^{-}\label{53} 
\end{eqnarray}
Now,

\begin{equation}
\label{54}
\left\langle \psi _{1}\right| (H-E_{0})\left| \psi _{1}\right\rangle =\left\langle \psi _{0}\right| U^{-1}(H-E_{0})U\left| \psi _{0}\right\rangle 
\end{equation}
The Hamiltonian H in (24) consists of two parts \( H_{z} \) and \( H_{XY} \)
. \( H_{z} \) commutes with U. Therefore,

\begin{eqnarray}
\left\langle \psi _{0}\right| U^{-1}HU\left| \psi _{0}\right\rangle  & = & \left\langle \psi _{0}\right| H_{z}\left| \psi _{0}\right\rangle +\left\langle \psi _{0}\right| U^{-1}H_{XY}U\left| \psi _{0}\right\rangle \nonumber \\
 & = & \left\langle \psi _{0}\right| H\left| \psi _{0}\right\rangle -\left\langle \psi _{0}\right| H_{XY}\left| \psi _{0}\right\rangle +\left\langle \psi _{0}\right| U^{-1}H_{XY}U\left| \psi _{0}\right\rangle \nonumber \\
 & = & E_{0}-\{\frac{1}{2}\left\langle S_{i}^{+}S_{i+1}^{_{-}}\right\rangle +h.c.\}+\left\langle \psi _{0}\right| U^{-1}H_{XY}U\left| \psi _{0}\right\rangle \label{55} 
\end{eqnarray}
In the last term, making use of the spin identities, we get 

\begin{eqnarray}
U^{-1}S^{+}_{i}S^{-}_{i+1}U & = & U^{-1}S^{+}_{i}UU^{-1}S^{-}_{i+1}U\nonumber \\
 & = & e^{-i\frac{2\pi }{L}i}S^{+}_{i}e^{i\frac{2\pi }{L}(i+1)}S^{-}_{i+1}\nonumber \\
 & = & e^{i\frac{2\pi }{L}}S^{+}_{i}S^{-}_{i+1}\label{56} 
\end{eqnarray}

So, from (54) - (56),

\begin{equation}
\label{57}
\left\langle \psi _{1}\right| (H-E_{0})\left| \psi _{1}\right\rangle =\frac{1}{2}\{\left[ e^{i\frac{2\pi }{L}}-1\right] \sum _{i}\left\langle S_{i}^{+}S_{i+1}^{_{-}}\right\rangle +h.c.\}
\end{equation}
Since the ground state is unique it must be isotropic implying 
\begin{equation}
\label{58}
\left\langle S_{i}^{+}S_{i+1}^{_{-}}\right\rangle =\left\langle S_{i}^{-}S_{i+1}^{+}\right\rangle 
\end{equation}
Thus,

\begin{equation}
\label{59}
\left\langle \psi _{1}\right| (H-E_{0})\left| \psi _{1}\right\rangle =\left[ cos(\frac{2\pi }{L})-1\right] \sum _{i}\left\langle S_{i}^{+}S_{i+1}^{_{-}}\right\rangle 
\end{equation}
Since \( S^{+}_{i}S^{-}_{i+1} \) is a bounded operator,

\begin{equation}
\label{60}
\left\langle \psi _{1}\right| (H-E_{0})\left| \psi _{1}\right\rangle =O(1/L)
\end{equation}
In the limit L \( \rightarrow \infty  \) , the r.h.s. of (60)\( \rightarrow  \)
0.

Constructing a low-energy state, however, proves nothing. The state may become
equal to the ground state as L \( \rightarrow \infty  \) . To complete the
proof, we have to show that this does not happen. This can be done by showing
that \( \left| \psi _{1}\right\rangle  \) are \( \left| \psi _{0}\right\rangle  \)
are orthogonal to each other. We will now show that \( \left| \psi _{1}\right\rangle  \)
has momentum \( \pi  \) relative to the ground state \( \left| \psi _{0}\right\rangle  \)
which , being unique, is also a momentum eigenstate. Thus the two states are
orthogonal to each other. Let T be the translation operator which produces a
translation by one site.

\begin{equation}
\label{61}
TS_{j}T^{-1}=S_{j+1},TS_{N}T^{-1}=S_{1}
\end{equation}

\begin{eqnarray}
TUT^{-1} & = & exp\left[ i\frac{2\pi }{L}\sum ^{L-1}_{n=1}nS_{n+1}^{z}+LS^{z}_{1}\right] \nonumber \\
 & = & Uexp\left[ -i\frac{2\pi }{L}\sum ^{L}_{n=1}S^{z}_{n}\right] e^{i2\pi S_{1}^{z}}\label{62} 
\end{eqnarray}
The ground state has total spin zero since it is unique and so the first exponential
has the value 1. Thus,

\begin{eqnarray}
T\left| \psi _{1}\right\rangle  & = & TU\left| \psi _{0}\right\rangle =TUT^{-_{1}}T\left| \psi _{0}\right\rangle \nonumber \\
 & = & e^{i2\pi S_{1}^{z}}UT\left| \psi _{0}\right\rangle \label{63} 
\end{eqnarray}
The exponential is -1 since \( S_{1}^{z} \) has eigenvalues \( \pm \frac{1}{2} \)
. So,

\begin{eqnarray}
T\left| \psi _{1}\right\rangle  & = & -UT\left| \psi _{0}\right\rangle =e^{i(\pi +k_{0})}U\left| \psi _{0}\right\rangle \nonumber \\
 & = & e^{i(\pi +k_{0})}\left| \psi _{1}\right\rangle \label{64} 
\end{eqnarray}
where \( k_{0} \) is the momentum wave vector of the ground state. The state
\( \left| \psi _{1}\right\rangle  \) has momentum wave vector \( \pi  \) relative
to the ground state \( \left| \psi _{0}\right\rangle  \) and is thus orthogonal
to it. This completes the proof of the LSM theorem for the S=1/2 Heisenberg
AFM Hamiltonian in 1d. The proof extends immediately to arbitrary half-odd integer
S but fails in the integer S case. For the latter, the exponential in (63) is
+1 and it cannot be proved that \( \left| \psi _{1}\right\rangle  \) and \( \left| \psi _{0}\right\rangle  \)
are orthogonal to each other. If the translational symmetry is spontaneously
broken in the ground state, the ground state develops degeneracy (Eq.(60) is
still true but \( \left| \psi _{1}\right\rangle  \) is another ground state
as L \( \rightarrow \infty  \) ). The excitation spectrum has a gap in this
case. 

In the presence of a magnetic field, the S=1/2 HAFM chain has gapless excitation
spectrum from zero field upto a saturation field, at which the ground state
becomes the fully polarized ferromagnetic ground state. For integer S, Haldane
made a conjecture that the excitation spectrum has a gap. This conjecture has
been verified both theoretically and experimentally \cite{10}. In the presence
of a magnetic field, the gap persists up to a critical field, equal to the gap.
The spectrum is gapless from the critical field to a saturation field. The Lieb-Scultz-Mattis
(LSM) theorem can be extended to the case of an applied magnetic field \cite{11}.
In this case, a striking phenomenon analogous to the quantum Hall effect occurs.
The phenomenon is that of magnetization plateaus. The content of the extended
theorem is : translationally invariant spin chains in an applied field can have
a gapped excitation spectrum, without breaking translational symmetry, only
when the magnetization/site m obeys the relation

\begin{equation}
\label{65}
S-m=integer
\end{equation}
where S is the magnitude of the spin. The proof is an easy extension of that
of the LSM theorem. The gapped phases correspond to magnetization plateaus in
the m vs. H curve at the quantized values of m which satisfy (65). Whenever
there is a gap in the spin excitation spectrum, it is obvious that the magnetization
cannot change in changing external field. Fractional quantization can also occur,
if accompanied by (explicit or spontaneous ) breaking of the translational symmetry.
In this case, the plateau condition is given by

\begin{equation}
\label{66}
n(S-m)=integer
\end{equation}
where n is the period of the ground state. Hida \cite{12} has considered a
S=1/2 HAFM chain with period 3 exchange coupling. A plateau in the magnetization
curve occurs at m=1/6 (1/3 of full magnetization). In this case, n = 3, S =
1/2 and m = 1/6 and the quantization condition in (66) is obeyed.\\
\textbf{D. Mermin-Wagner's theorem \cite{6,13}} :\\
There cannot be any AFM LRO at finite T in dimensions d = 1 and 2. AFM LRO can,
however, exist in the ground states of spin models in d = 2. LRO exists in the
ground state of the 3d HAFM model for spin S\( \geq  \) 1/2 \cite{14}. At
finite T, the LRO persists upto a critical temperature \( T_{c} \). For square
\cite{15} and hexagonal \cite{16} lattices, LRO exists in the ground state
for S\( \geq  \) 1. The above results are based on rigorous proofs. No such
proof exists as yet for S=1/2, d =2 ( this case is of interest because the \( CuO_{2} \)
plane of the high-\( T_{c} \) cuprate systems is a S = 1/2 2d AFM ). \\
\textbf{3. Possible ground states and excitation spectra}

The Néel state is not the ground state of the quantum Heisenberg Hamiltonian.
the Néel state, we know, has perfect AFM LRO. It is still possible that the
true ground state of a quantum Hamiltonian has Néel-type order, i.e., non-zero
sublattice magnetization. The ground state is essentially the Néel state with
quantum fluctuations mixed in. This picture does indeed describe the ground
state of the HAFM in d \( \geq  \) 2. The ground state of the HAFM in 1d has
no LRO and is an example of a spin liquid . Even in higher dimensions, one can
have disordered ground states due to what is known as frustration. Frustration
occurs when the interaction energies associated with all the spin pairs cannot
be simultaneously minimized. This may occur due to lattice topology or due to
the presence of further-neighbour interactions. The triangular \cite{17}, kagomé\cite{18}
and pyrochlore lattices \cite{19} are well-known examples of frustrated lattices.
Another recent example is the pentagonal lattice \cite{20}. In these lattices,
the number of bonds in each elementary plaquette is odd. We illustrate frustration
by considering Ising spins on a triangular plaquette. Each Ising spin, sitting
at a vertex of the triangle, can be either up or down. AFM exchange interaction
energy is minimized if the n.n. spin pairs are antiparallel. As Figure 1 shows,
not all the three n.n. spin pairs can be simultaneously antiparallel. The parallel
spin pair may be located on any one of the bonds of the triangle. In fact, the
ground state of the Ising AFM on the triangular lattice is highly degenerate.
The entropy/spin has a non-zero value in the ground state and there is no ordering
of the spins. If there is a mixture of FM and AFM interactions, then a spin
model defined on an unfrustrated lattice like the square lattice (each elementary
plaquette, a square, has four bonds) can also exhibit frustration, if the number
of n.n. AFM interactions in a plaquette is odd. Frustration also occurs if both
n.n. as well as n.n.n. interactions are present. Frustration favours disordered
ground states but does not rule out AFM LRO in the ground state. The Ising AFM
on the triangular lattice has a disordered ground state but various calculations
suggest that the HAFM Hamiltonian on the triangular lattice has AFM LRO \cite{21}.

Various types of disordered ground states are possible: dimer, resonating valence
bond (RVB), twisted, chiral and strip or collinear states. The lowest energy
configuration of two spins interacting via the HAFM exchange interaction is
a spin singlet with energy -3J/4. The singlet is often termed a valence bond
(VB) or dimer (an object which occupies two sites). A pictorial representation
of a VB between spins at sites i and j is a solid line joining the sites. The
square lattice with N sites can be covered by N/2 VBs in various possible ways.
Each possibility corresponds to a VB state. A RVB state is a coherent linear
superposition of VB states (Figure 2). The 2d S = 1/2 AFM \( CaV_{4}O_{9} \)
has a lattice structure corresponding to a 1/5-depleted square lattice. The
lattice consists of plaquettes of four spins connected by bonds. The ground
state of the AFM is the Plaquette RVB (PRVB) state {[}22{]}. In this state,
the spin configuration in each plaquette is a linear combination of two VB configurations.
In one of these the two singlet bonds are horizontal and in the other they are
vertical. Bose and Ghosh {[}23{]} have constructed a spin model for which the
PRVB state is the exact ground state. The triplet excitation is separated from
the ground state by an energy gap and singlet excitations fall within the gap.

In a columnar dimer state, the dimers or VBs are arranged in columns. A description
of the other disordered ground states can be obtained from Ref. {[}24{]}. The
disordered ground states do not have conventional LRO but are not simple paramagnets.
One can define suitable operators, the expectation values of which in the appropriate
ground states are non-zero. The states, with different types of order, are described
as quantum paramagnets. There is a huge literature on AFM spin models with disordered
ground states. This is mainly because of the relevance of such studies for high
temperature superconductivity. The common structural ingredient of the high-\( T_{c} \)
cuprate systems is the \( CuO_{2} \) plane. It looks like a square lattice
with \( Cu^{2+} \) ions located at the lattice sites and oxygen ions sitting
on the bonds in between. The \( Cu^{2+} \) ions carry spin-1/2 and the n.n.
spins interact via the HAFM exchange interaction Hamiltonian. In the undoped
state, the cuprates exhibit AFM LRO below the Néel temperature. On doping the
systems with a few percent of holes, the AFM LRO is rapidly destroyed leaving
behind a spin disordered state. It is important to understand the nature of
the spin disordered state as this state serves as the reference state into which
holes are introduced on doping. The holes are responsible for charge transport
in the metallic state and forms bound pairs in the superconducting state. We
now discuss some spin models for which the ground state properties (energy and
wave function as well as correlation functions) can be calculated exactly. The
first model is described by the Majumdar-Ghosh Hamiltonian \cite{1,6}. This
is defined in 1d for spins of magnitude 1/2. The Hamiltonian is AFM in nature
and includes both n.n. as well as n.n.n. interactions. The strength of the latter
is half that of the former.

\begin{equation}
\label{67}
H_{MG}=J\sum ^{N}_{i=1}\overrightarrow{S}_{i}.\overrightarrow{S}_{i+1}+J/2\sum ^{N}_{i=1}\overrightarrow{S}_{i}.\overrightarrow{S}_{i+2}
\end{equation}
The exact ground state of \( H_{MG} \) is doubly degenerate and the states
are

\begin{eqnarray}
\phi _{1} & \equiv  & [12][34].......[N-1N]\nonumber \\
\phi _{2} & \equiv  & [23][45]........[N1]\label{68} 
\end{eqnarray}
where {[}lm{]} denotes a singlet spin configuration for spins located at the
sites l and m. Also, PBC is assumed. One finds that translational symmetry is
broken in the ground state. The proof that \( \phi _{1} \) and \( \phi _{2} \)
are the exact ground states can be obtained by the method of `divide and conquer'.
One can verify that \( \phi _{1} \) and \( \phi _{2} \) are exact eigenstates
of \( H_{MG} \) by applying the spin identity

\begin{equation}
\label{69}
\overrightarrow{S}_{n}.(\overrightarrow{S}_{l}+\overrightarrow{S}_{m})[lm]=0
\end{equation}
Let \( E_{1} \) be the energy of \( \phi _{1} \) and \( \phi _{2} \) . Let
\( E_{g} \) be the exact ground state energy. Then

\begin{equation}
\label{70}
E_{g}\leq E_{1}
\end{equation}
One divides the Hamiltonian \( H \) into sub-Hamiltonians, \( H_{i} \) 's,
such that \( H=\sum _{i}H_{i}. \) \( H_{i} \) can be exactly diagonalised
and let \( E_{i0} \) be the ground state energy. Let \( \psi _{g} \) be the
exact ground state wave function. By variational theory, 

\begin{equation}
\label{71}
E_{g}=\left\langle \psi _{g}\right| H\left| \psi _{g}\right\rangle =\sum _{i}\left\langle \psi _{g}\right| H_{i}\left| \psi _{g}\right\rangle \geq \sum _{i}E_{i0}
\end{equation}
 From (70) and (71), one gets

\begin{equation}
\label{72}
\sum _{i}E_{i0}\leq E_{g}\leq E_{1}
\end{equation}
 If one can show that \( \sum _{i}E_{i0}=E_{1} \) , then \( E_{1} \) is the
exact ground state energy. For the MG-chain, the sub-Hamiltonian \( H_{i} \)
is

\begin{equation}
\label{73}
H_{i}=J/2(\overrightarrow{S}_{i}.\overrightarrow{S}_{i+1}+\overrightarrow{S}_{i+1}.\overrightarrow{S}_{i+2}+\overrightarrow{S}_{i+2}.\overrightarrow{S}_{i})
\end{equation}
There are N such sub-Hamiltonians. One can easily verify that \( E_{i0}=-3J/8 \)
and \( E_{1}=-3\frac{J}{4}\frac{N}{2} \) (-3J/4 is the energy of a singlet
and there are N/2 VBs in \( \phi _{1} \) and \( \phi _{2} \) ). From (72),
one finds that the lower and upper bounds of \( E_{g} \) are equal and hence
\( \phi _{1} \) and \( \phi _{2} \) are the exact ground states with energy
\( E_{1}=-3JN/8 \) . There is no LRO in the two-spin correlation function in
the ground state:

\begin{equation}
\label{74}
K^{2}(i,j)=\left\langle S_{i}^{z}S_{j}^{z}\right\rangle =\frac{1}{4}\delta _{ij}-\frac{1}{8}\delta _{\mid i-j\mid ,1}
\end{equation}
 The four-spin correlation function has off-diagonal LRO.

\begin{eqnarray}
K^{4}(ij,lm) & = & \left\langle S_{i}^{x}S_{j}^{x}S_{l}^{y}S_{m}^{y}\right\rangle \nonumber \\
 & = & K^{2}(ij)K^{2}(lm)+\frac{1}{64}\delta _{\mid i-j\mid ,1}\delta _{\mid l-m\mid ,1}\nonumber \\
 & \times  & exp(i\pi (\frac{i+j}{2}-\frac{l+m}{2}))\label{75} 
\end{eqnarray}
Let T be the translation operator for unit displacement. Then

\begin{equation}
\label{76}
T\phi _{1}=\phi _{2},T\phi _{2}=\phi _{1}
\end{equation}
 The states

\begin{equation}
\label{77}
\phi ^{+}=\frac{1}{\sqrt{2}}(\phi _{1}+\phi _{2}),\phi ^{-}=\frac{1}{\sqrt{2}}(\phi _{1}-\phi _{2})
\end{equation}
correspond to momentum wave vectors k = 0 and k = \( \pi  \) . The excitation
spectrum is not exactly known. Shastry and Sutherland {[}25{]} have derived
the excitation spectrum in the basis of `defect' states. A defect state has
the wave function

\begin{eqnarray}
\psi (p,m) & = & ....[2p-3,2p-2]\alpha _{2p-1}[2p,2p+1]\nonumber \\
 &  & ....[2m-2,2m-1]\alpha _{2m}[2m+1,2m+2]\label{78} 
\end{eqnarray}
where the defects (\( \alpha _{2p-1} \) and \( \alpha _{2m} \) ) separate
two ground states. The two defects are up-spins and the total spin of the state
is 1. Similarly, the defect spins can be in a singlet spin configuration so
that the total spin of the state is zero. Because of PBC, the defects occur
in pairs. A variational state can be constructed by taking a linear combination
of the defect states. The excitation spectrum consists of a continuum with a
lower edge at \( J(5/2-2\mid cosk\mid ) \). A bound state of the two defects
can occur in a restricted region of momentum wave vectors. The MG chain has
been studied for general values \( \alpha J \) of the n.n.n. interaction {[}26{]}.
The ground state is known exactly only at the MG point \( \alpha =1/2 \) .The
excitation spectrum is gapless for 0 < \( \alpha  \) < \( \alpha _{cr} \)
(\( \simeq  \) 0.2411).Generalizations of the MG model to two dimensions exist
{[}27,28{]}. The Shastry-Sutherland model {[}27{]} is defined on a square lattice
and includes diagonal interactions as shown in Figure 3. The n.n. and diagonal
exchange interactions are of strength \( J_{1} \) and \( J_{2} \) respectively.
For \( \frac{J_{1}}{J_{2}} \) below a critical value \( \sim  \) 0.7, the
exact ground state consists of singlets along the diagonals. At the critical
point, the ground state changes from the gapful disordered state to the AFM
ordered gapless state. The compound \( SrCu_{2}(BO_{3})_{2} \) is well-described
by the Shastry-Sutherland model {[}29{]}. Bose and Mitra {[}28{]} have constructed
a \( J_{1}-J_{2}-J_{3}-J_{4}-J_{5} \) spin-1/2 model on the square lattice.
\( J_{1},J_{2},J_{3},J_{4} \) and \( J_{5} \) are the strengths of the n.n.,
diagonal , n.n.n., knight's-move-distance-away and further-neighbour -diagonal
exchange interactions (Figure 4). The four columnar dimer states (Figure 5)
have been found to be the exact eigenstates of the spin Hamiltonian for the
ratio of interaction strengths 

\begin{equation}
\label{79}
J_{1}:J_{2}:J_{3}:J_{4}:J_{5}=1:1:1/2:1/2:1/4
\end{equation}
 It has not been possible as yet to prove that the four columnar dimer states
are also the ground states. Using the method of `divide and conquer', one can
only prove that a single dimer state is the exact ground state with the dimer
bonds of strength 7J. The strengths of the other exchange interactions are as
specified in (79).

In Section 2, we have discussed the LSM theorem, the proof of which fails for
integer spin chains. Haldane {[}30,10{]} in 1983 made the conjecture, based
on a mapping of the HAFM Hamiltonian, in the long wavelength limit, onto the
nonlinear \( \sigma  \) model, that integer-spin HAFM chains have a gap in
the excitation spectrum. The conjecture has now been verified both theoretically
and experimentally {[}31{]}. In 1987, Affleck, Kennedy, Lieb and Tasaki (AKLT)
constructed a spin-1 model in 1d for which the ground state could be determined
exactly {[}32{]}. Consider a 1d lattice, each site of which is occupied by a
spin-1. Each such spin can be considered to be a symmetric combination of two
spin-1/2's. Thus, one can write down

\begin{eqnarray}
\psi _{++} & = & \left| ++\right\rangle ,S^{z}=+1\nonumber \\
\psi _{--} & = & \left| --\right\rangle ,S^{z}=-1\nonumber \\
\psi _{+-} & = & \frac{1}{\sqrt{2}}(\left| +-\right\rangle +\left| -+\right\rangle ),S^{z}=0\nonumber \\
\psi _{-+} & = & \psi _{+-}\label{80} 
\end{eqnarray}
where `\( + \)' (`\( - \)') denotes an up (down) spin.

AKLT constructed a valence bond solid (VBS) state in the following manner. In
this state, each spin-1/2 component of a spin-1 forms a singlet (valence bond)
with a spin-1/2 at a neighbouring site. Let \( \epsilon ^{\alpha \beta } \)
(\( \alpha ,\beta =+ \)or \( - \)) be the antisymmetric tensor :

\begin{equation}
\label{81}
\epsilon ^{++}=\epsilon ^{--}=0,\epsilon ^{+-}=-\epsilon ^{-+}=1
\end{equation}
A singlet spin configuration can be expressed as \( \frac{1}{\sqrt{2}}\epsilon ^{\alpha \beta }\left| \alpha \beta \right\rangle  \)
, summation over repeated indices being implied. The VBS wave function (with
PBC) can be written as

\begin{eqnarray}
\left| \psi _{VBS}\right\rangle  & = & 2^{-\frac{N}{2}}\psi _{\alpha _{1}\beta _{1}}\epsilon ^{\beta _{1}\alpha _{2}}\psi _{\alpha _{2}\beta _{2}}\epsilon ^{\beta _{2}\alpha _{3}}.......\nonumber \\
 &  & \psi _{\alpha _{i}\beta _{i}}\epsilon ^{\beta _{i}\alpha _{i+1}}........\psi _{\alpha _{N}\beta _{N}}\epsilon ^{\beta _{N}\alpha _{1}}\label{82} 
\end{eqnarray}
\( \left| \psi _{VBS}\right\rangle  \) is a linear superposition of all configurations
in which each \( S^{z}=+1 \) is followed by a \( S^{z}=-1 \) with an arbitrary
number of \( S^{z}=0 \) spins in between and vice versa. If one leaves out
the zero's, one gets a Néel-type of order. One can define a non-local string
operator

\begin{equation}
\label{83}
\sigma ^{\alpha }_{ij}=-S^{\alpha }_{i}exp(i\pi \sum ^{j-1}_{l=i+1}S^{\alpha }_{l})S^{\alpha }_{j},(\alpha =x,y,z)
\end{equation}
and the order parameter

\begin{equation}
\label{84}
O^{\alpha }_{string}=lim_{\mid i-j\mid \rightarrow \infty }\left\langle \sigma ^{\alpha }_{ij}\right\rangle 
\end{equation}
 The VBS state has no conventional LRO but is characterised by a non-zero value
4/9 of \( O^{\alpha }_{string} \). After constructing the VBS state, AKLT determined
the Hamiltonian for which the VBS state is the exact ground state. The Hamiltonian
is

\begin{equation}
\label{85}
H_{AKLT}=\sum _{i}P_{2}(\overrightarrow{S_{i}}+\overrightarrow{S}_{i+1})
\end{equation}
where \( P_{2} \) is the projection operator onto spin 2 for a pair of n.n.
spins. The presence of a VB between each neighbouring pair implies that the
total spin of each pair cannot be 2 (after two of the S=1/2 variables form a
singlet, the remaining S=1/2's could form either a triplet or a singlet). Thus,
\( H_{AKLT} \) acting on \( \left| \psi _{VBS}\right\rangle  \) gives zero.
Since \( H_{AKLT} \) is a sum over projection operators, the lowest possible
eigenvalue is zero. Hence, \( \left| \psi _{VBS}\right\rangle  \) is the ground
state of \( H_{AKLT} \) with eigenvalue zero. The AKLT ground state (the VBS
state) is spin-disordered and the two-spin correlation function has an exponential
decay.The total spin of two spin-1's is 2, 1 or 0. The projection operator onto
spin j for a pair of n.n. spins has the general form

\begin{equation}
\label{86}
P_{j}(\overrightarrow{S}_{i}+\overrightarrow{S}_{i+1})=\prod _{l\neq j}\frac{\left[ l(l+1)-\overrightarrow{S}^{2}\right] }{\left[ l(l+1)-j(j+1)\right] }
\end{equation}
where \( \overrightarrow{S}=\overrightarrow{S_{i}}+\overrightarrow{S}_{i+1}. \)
For the AKLT model, j =2 and l = 1, 0. From (85) and (86),

\begin{equation}
\label{87}
H_{AKLT}=\sum _{i}\left[ \frac{1}{2}(\overrightarrow{S}_{i}.\overrightarrow{S}_{i+1})+\frac{1}{6}(\overrightarrow{S}_{i}.\overrightarrow{S}_{i+1})^{2}+\frac{1}{3}\right] 
\end{equation}
The method of construction of the AKLT Hamiltonian can be extended to higher
spins and to dimensions d >1. The MG Hamiltonian (apart from a numerical prefactor
and a constant term) can be written as

\begin{equation}
\label{88}
H=\sum _{i}P_{3/2}(\overrightarrow{S}_{i}+\overrightarrow{S}_{i+1}+\overrightarrow{S}_{i+2})
\end{equation}
 The S=1 HAFM and the AKLT chains are in the same Haldane phase, characterised
by a gap in the excitation spectrum. The physical picture provided by the VBS
ground state of the AKLT Hamiltonian holds true for real systems {[}33{]}. The
excitation spectrum of \( H_{AKLT} \) cannot be determined exactly. Arovas
et al {[}34{]} have proposed a trial wave function 

\begin{equation}
\label{89}
\left| k\right\rangle =N^{-\frac{1}{2}}\sum ^{N}_{j=1}e^{ikj}S^{\mu }_{j}\left| \psi _{VBS}\right\rangle ,\mu =z,+,-
\end{equation}
and obtained

\begin{equation}
\label{90}
\epsilon (k)=\frac{\left\langle k\right| H_{VBS}\left| k\right\rangle }{<k\mid k>}=\frac{25+15cos(k)}{27}
\end{equation}
The gap in the excitation spectrum \( \Delta  \) = \( \frac{10}{27} \) at
k = \( \pi  \). Another equivalent way of creating excitations is to replace
a singlet bond by a triplet spin configuration {[}35{]}. 

We end this Section with a brief mention of spin ladders. The subject of spin
ladders is a rapidly growing area of research {[}36,37,38{]}. Ref. {[}38{]}
gives an exhaustive overview of real systems with ladder-like structure and
the results of various experiments on ladder compounds. A spin ladder consists
of n chains ( n = 2,3,4,...etc.) coupled by rungs. The simplest spin ladder
(Figure 6) corresponds to n = 2. Each site of the ladder is occupied by a spin-1/2.
Let J and \( J_{R} \) be the strengths of the intrachain n.n. and rung exchange
interactions. When \( J_{R}=0 \) , the ladder decouples into two HAFM chains
for which the excitation spectrum is known to be gapless. A gap opens up in
the excitation spectrum even for an infinitesimal value of \( J_{R} \) and
persists for all values of \( J_{R} \) . When \( J_{R} \) is >\textcompwordmark{}>
J, the ground state predominantly consists of singlets along the rungs of the
ladder. The ground state is spin-disordered. A triplet excitation is created
by replacing a rung singlet by a triplet and letting it propagate. The n-chain
spin ladder exhibits the `odd-even' effect. The excitation spectrum is gapless
(with gap) when n is odd (even). When n is odd, the spin-spin correlation function
has a power-law decay as in the case of the HAFM S = 1/2 chain. For n even,
the ground state correlation function has an exponential decay. For n sufficiently
large, the square lattice limit is reached with AFM LRO in the ground state
and a gapless excitation spectrum. Bose and Gayen {[}39{]} have constructed
a two-chain spin ladder model which includes diagonal exchange interactions
of strength J. When \( J_{R} \) >\textcompwordmark{}> 2J, the exact ground
state consists of singlets along the rungs. Later, Xian {[}39{]} has shown that
the same exact ground state is obtained for \( \frac{J_{R}}{J}>\frac{J_{R}}{J_{c}}\simeq  \)
1.40148. Ghosh and Bose {[}40{]} have generalised the two-chain ladder model
to an n-chain model for which the `odd-even' effect can be exactly demonstrated. 

The spin ladder can be doped with holes. Two holes can form a bound state as
in the case of high-\( T_{c} \) cuprate systems. The binding of two holes in
a t-J ladder model has been shown through exact, analytic calculations {[}41{]}.
A ladder compound has been discovered which shows superconductivity (maximum
\( T_{c} \) \( \sim  \) 12K ) on doping with holes and under pressure. In
the superconducting state, the holes form bound pairs. Again, the analogy with
the cuprate systems is strong. For detailed information on undoped and doped
spin ladders, see the Refs. {[}36,38{]}.\\
4. The Bethe Ansatz

The Bethe Ansatz (BA) was formulated by Bethe in 1931 {[}42{]} and describes
a wave function with a particular kind of structure. Bethe considered a well-known
model in magnetism, the spin-1/2 Heisenberg linear chain in which only n.n.
spins interact. The wave function of the interacting many body system has the
BA form. For ferromagnetic (FM) interactions, Bethe starting with the BA wave
function derived the energy dispersion relations of spin wave (magnon) bound
states exactly. Hulthén {[}43{]} used the BA method to derive an exact expression
for the ground state energy of the antiferromagnetic (AFM) spin-1/2 Hamiltonian.
In later years, it was realized that the method has a wider applicability and
is not confined to magnetic spin chains. In fact, many of the exact solutions
of interacting many body systems are but variations or generalizations of Bethe's
method. Examples include the Fermi and Bose gas models in which particles on
a line interact through delta function potentials {[}44{]}, the Hubbard model
in 1d {[}45{]}, 1d plasma which crystallizes as a Wigner solid {[}46{]}, the
Lai-Sutherland model {[}47{]}which includes the Hubbard model and a dilute magnetic
model as special cases, the Kondo model in 1d {[}48{]}, the single impurity
Anderson model in 1d {[}49{]}, the supersymmetric t-J model (J =2t) {[}50{]}
etc. In the case of quantum models, the BA method is applicable only to 1d models.
The BA method has also been applied to derive exact results for classical lattice
statistical models in 2d. In the following we show how the BA method works by
considering the spin-1/2 linear chain Heisenberg model mentioned in the beginning.
The Hamiltonian for the linear chain is given by

\begin{equation}
\label{91}
H=-J\sum ^{N}_{i=1}\overrightarrow{S}_{i}.\overrightarrow{S}_{i+1}
\end{equation}
 describing the interaction of spins located at n.n. lattice sites. The total
number of spins is given by N. The magnitude of the spin at each site is given
by 1/2. Periodic boundary condition (PBC) is assumed so that N + 1 \( \equiv  \)1.
We consider the interaction to be ferromagnetic so that the exchange constant
J is > 0. The Hamiltonian describes magnetic insulators in which magnetic moments
are localized on well-separated atoms. The Hamiltonian (91) can be written as

\begin{equation}
\label{92}
H=-J\sum ^{N}_{i=1}\left[ S^{z}_{i}S^{z}_{i+1}+\frac{1}{2}(S^{+}_{i}S^{-}_{i+1}+S^{-}_{i}S^{+}_{i+1})\right] 
\end{equation}
There are two conserved quantities:

\begin{eqnarray}
S^{z} & = & \sum ^{N}_{i=1}S^{z}_{i},[S^{z},H]=0\nonumber \\
\overrightarrow{S} & = & \sum ^{N}_{i=1}\overrightarrow{S}_{i},[\overrightarrow{S}^{2},H]=0\label{93} 
\end{eqnarray}
Since \( S^{z} \) is a conserved quantity, there is no mixing of subspaces
corresponding to the different values of \( S^{z} \), thus making it easier
to solve the eigenvalue problem. Define the spin deviation operator

\begin{equation}
\label{94}
n=\frac{N}{2}+S^{z}
\end{equation}
 The FM ground state \( \psi _{g} \) corresponds to n =0, i.e., \( S^{z}=-\frac{N}{2} \)
with all the spins pointing downwards:

\begin{equation}
\label{95}
\psi _{g}=\beta (1)\beta (2).......\beta (N)
\end{equation}
 The ground state energy is \( E_{g}=-J\frac{N}{4} \) .

For n = r, we have r up -spins at the sites \( m_{1},m_{2},.....m_{r} \). The
eigenfunction \( \psi  \) of H in the n = r subspace is a linear combination
of the \( ^{N}C_{r} \) functions \( \psi (m_{1},m_{2},......,m_{r}) \) :

\begin{equation}
\label{96}
\psi =\sum _{\{m\}}a(m_{1},m_{2},......,m_{r})\psi (m_{1},m_{2},......,m_{r}),\{m\}=(m_{1},m_{2},......,m_{r})
\end{equation}
The summation is over all the \( m_{i} \) 's running over 1 to N subject to
the condition

\begin{equation}
\label{97}
m_{1}<m_{2}<.......<m_{r}
\end{equation}
to avoid overcounting of states. We now consider the operation of the Hamiltonian
H given in (92) on \( \psi  \). The rules of the spin algebra are:

(a) if two neighbouring spins are parallel, the term \( S^{+}S^{-}+S^{-}S^{+} \)
in H gives no contribution. The \( S^{z}S^{z} \) term on the other hand gives
\( -\frac{J}{4} \) times the same state \( \psi (\{m\}) \) ,

(b) for an antiparallel n.n. spin pair, the term \( S^{+}S^{-}+S^{-}S^{+} \)
interchanges the spins in \( \psi (\{m\}) \) with a multiplicative constant
\( -\frac{J}{2} \) . The \( S^{z}S^{z} \) term gives a contribution \( \frac{J}{4} \)
multiplied by the same state \( \psi (\{m\}) \) . The eigenvalue equation is
given by

\begin{equation}
\label{98}
H\psi =E\psi =E\sum _{\{m\}}a(\{m\})\psi (\{m\})
\end{equation}
 Take the scalar product with a particular \( \psi ^{*}(m_{1},m_{2},......,m_{r}) \)
and use orthogonality properties to get

\begin{equation}
\label{99}
Ea(m_{1},m_{2},......,m_{r})=-\frac{J}{2}\sum _{\{m^{\prime }\}}a(\{m^{\prime }\})+\frac{J}{4}N^{\prime }a(\{m\})-\frac{J}{4}(N-N^{\prime })a(\{m\})
\end{equation}
 \( N^{\prime } \) is the total number of antiparallel spin pairs so that (\( N \)
- \( N^{\prime } \) ) is the number of parallel spin pairs. The sum is over
all \( N^{\prime } \) distributions \( (m_{1}^{\prime },m_{2}^{\prime },......,m_{r}^{\prime }) \)
which arise from a n.n. exchange of antiparallel spins in \( (m_{1},m_{2},......,m_{r}) \).
Eq. (99) can further be written as

\begin{equation}
\label{100}
(E+JN)a(m_{1},m_{2},......,m_{r})=-\frac{J}{2}\sum _{\{m^{\prime }\}}\left[ a(m_{1}^{\prime },m_{2}^{\prime },......,m_{r}^{\prime })-a(m_{1},m_{2},......,m_{r})\right] 
\end{equation}
Or

\begin{equation}
\label{101}
2\epsilon a(m_{1},m_{2},......,m_{r})=-\sum _{\{m^{\prime }\}}\left[ a(m_{1}^{\prime },m_{2}^{\prime },......,m_{r}^{\prime })-a(m_{1},m_{2},......,m_{r})\right] 
\end{equation}
where

\begin{equation}
\label{102}
\epsilon =\frac{1}{J}(E+\frac{N}{4}J)
\end{equation}
 is the energy, in units of J, of the excited states w.r.t. that of the ground
state energy \( E_{g}. \) Next, one considers the PBC from which one gets

\begin{equation}
\label{103}
a(m_{1},....,m_{i},.....,m_{r})=a(m_{1},......,m_{r},m_{i}+N)
\end{equation}
 The spin waves of a FM system are obtained for n =1.

\begin{equation}
\label{104}
\psi (m)=\sum _{m}a(m)\psi (m)
\end{equation}
 
\begin{equation}
\label{105}
2\epsilon a(m)=-\left[ a(m+1)+a(m-1)-2a(m)\right] 
\end{equation}
The solution of (105) is given by 

\[
a(m)=e^{ikm}\]
and

\begin{equation}
\label{106}
\epsilon =1-cosk
\end{equation}
 From PBC, the allowed values of k are 

\begin{equation}
\label{107}
k=\frac{2\pi \lambda }{N},\lambda =0,1,...,N-1
\end{equation}
 There are N states. Consider now the case n =2, i.e., there are two spin deviations.
We have to take into account two cases. If the two spin deviations are not neighbours,
one gets

\begin{eqnarray}
2\epsilon a(m_{1},m_{2}) & = & -a(m_{1}+1.m_{2})-a(m_{1}-1,m_{2})-a(m_{1},m_{2}+1)\nonumber \\
 &  & -a(m_{1},m_{2}-1)+4a(m_{1},m_{2})\label{108} 
\end{eqnarray}
 If the spin deviations are neighbours, one gets

\begin{equation}
\label{109}
2\epsilon a(m_{1},m_{1}+1)=-a(m_{1}-1,m_{1}+1)-a(m_{1},m_{1}+2)+2a(m_{1},m_{1}+1)
\end{equation}
 Eq. (108) is satisfied by the Ansatz

\begin{equation}
\label{110}
a(m_{1},m_{2})=C_{1}e^{i(k_{1}m_{1}+k_{2}m_{2})}+C_{2}e^{i(k_{2}m_{1}+k_{1}m_{2})}
\end{equation}
Then

\begin{equation}
\label{111}
\epsilon =1-cosk_{1}+1-cosk_{2}
\end{equation}
 \( C_{1},C_{2},k_{1} \) and \( k_{2} \) are to be determined. Eq. (109) can
be satisfied by Eq. (110), if the coefficients \( C_{1} \) and \( C_{2} \)
are chosen to make

\begin{equation}
\label{112}
a(m_{1},m_{1})+a(m_{1}+1,m_{1}+1)-2a(m_{1},m_{1}+1)=0
\end{equation}
 For spin-1/2 particles, the amplitudes a(m,m) have no physical significance
but are defined by Eq. (112). Put Eq. (110) into Eq.(112) to get

\begin{equation}
\label{113}
\frac{C_{1}}{C_{2}}=\frac{sin\frac{1}{2}(k_{1}-k_{2})+i\{cos\frac{1}{2}(k_{1}+k_{2})-cos\frac{1}{2}(k_{1}-k_{2})\}}{sin\frac{1}{2}(k_{1}-k_{2})-i\{cos\frac{1}{2}(k_{1}+k_{2})-cos\frac{1}{2}(k_{1}-k_{2})\}}
\end{equation}
Let \( C_{1}=e^{i\frac{\phi }{2}} \) and \( C_{2}=e^{-i\frac{\phi }{2}} \).
Then

\begin{equation}
\label{114}
2cot\frac{\phi }{2}=cot\frac{k_{1}}{2}-cot\frac{k_{2}}{2}
\end{equation}
 Also, PBC gives

\begin{equation}
\label{115}
a(m_{1},m_{2})=a(m_{2},m_{1}+N)
\end{equation}
 This is in accordance with the ordering of \( m_{i} \) 's given in Eq. (97).
From the PBC, one gets

\begin{eqnarray}
Nk_{1}-\phi  & = & 2\pi \lambda _{1}\nonumber \\
Nk_{2}+\phi  & = & 2\pi \lambda _{2}\nonumber \\
\lambda _{1},\lambda _{2} & = & 0,1,2,.....,(N-1)\label{116} 
\end{eqnarray}
 We have three equations (114) and (116) for the three quantities \( \phi _{1},k_{1},k_{2} \)
. The sum of \( k_{1} \) and \( k_{2} \) is a constant of motion by translational
symmetry.

\begin{equation}
\label{117}
k=k_{1}+k_{2}=\frac{2\pi }{N}(\lambda _{1}+\lambda _{2})
\end{equation}
Since \( \lambda _{1} \) and \( \lambda _{2} \) can be interchanged without
affecting the solution, we choose \( \lambda _{1}\leq \lambda _{2} \). For
a given \( \lambda _{1} \) and \( \lambda _{2} \), the solutions are completely
determined by (114) and (116). 

The following three cases are to be considered for determining the dependence
of the wave numbers \( k_{1} \) and \( k_{2} \) on \( \lambda _{1} \) and
\( \lambda _{2} \) :\\
Case I: \( \lambda _{2}\geq \lambda _{1}+2 \)\\
Case II: \( \lambda _{1}=\lambda _{2}=\frac{\lambda }{2} \) , \( \lambda  \)
even\\
Case III: \( \lambda _{1}=\lambda _{2}-1=\frac{(\lambda -1)}{2} \), \( \lambda  \)
odd\\
In Case I, for a given value of \( \lambda _{2} \), \( \lambda _{1} \) can
take on the values \( \lambda _{1} \) = 0,1,2,....,\( \lambda _{2} \) -2.
The total number of solutions is given by 

\begin{equation}
\label{118}
\sum ^{N-1}_{\lambda _{2}=2}(\lambda _{2}-1)=^{N-1}C_{2}
\end{equation}
 In this case, all the wave numbers are real and for specified values of \( \lambda _{1} \)
and \( \lambda _{2} \) , \( k_{1},k_{2} \) and \( \phi  \) are determined
uniquely from (114) and (116). The energy eigenvalue \( \epsilon _{I} \) is
given by (111) with \( k_{1},k_{2} \) real. In Case II, from (116),

\begin{eqnarray}
Nk_{1}-\phi  & = & \pi \lambda \nonumber \\
Nk_{2}+\phi  & = & \pi \lambda \label{119} 
\end{eqnarray}
We look for solutions of the form

\begin{eqnarray}
k_{1} & = & u+iv\nonumber \\
k_{2} & = & u-iv\nonumber \\
\phi  & = & \psi +i\chi \label{120} 
\end{eqnarray}
From (119) and (120), we find that \( \psi  \) = 0 and \( \chi =Nv \) . If
\( v \) is non-zero, then for \( N\rightarrow \infty  \), \( \chi  \) also
\( \rightarrow \infty  \). So

\begin{eqnarray*}
cot\frac{\phi }{2} & = & \frac{sin\psi -isinh\chi }{cosh\chi -cos\psi }\\
 & = & -i(1+2e^{-\chi })
\end{eqnarray*}
In a first approximation,

\begin{eqnarray}
2cot\frac{\phi }{2} & = & cot\frac{k_{1}}{2}-cot\frac{k_{2}}{2}\nonumber \\
\Rightarrow -2i & = & \frac{sinu-isinhv}{coshv-cosu}-\frac{sinu+isinhv}{coshv-cosu}\nonumber \\
or,sinhv & = & coshv-cosu\nonumber \\
or,e^{-v} & = & cosu\label{121} 
\end{eqnarray}
Hence, 
\begin{eqnarray}
\epsilon  & = & 2-cos(u+iv)-cos(u-iv)\nonumber \\
 & = & 2-cosu(cosu+\frac{1}{cosu})\nonumber \\
 & = & \frac{1}{2}(1-cos2u)=\frac{1}{2}(1-cosk)\label{122} 
\end{eqnarray}
From (121), cosu\( \geq  \)0, so \( -\pi /2\leq u\leq \pi /2 \) and k can
be chosen to be in the range \( -\pi \leq k\leq \pi  \). Let us specify the
energy obtained in (122), for complex values of \( k_{1} \) and \( k_{2} \)
, as \( \epsilon _{II} \) . Let us now compare \( \epsilon _{I} \) and \( \epsilon _{II} \)
. From (111),

\begin{eqnarray}
\epsilon _{I} & = & 1-cosk_{1}+1-cosk_{2}\nonumber \\
 & = & 2-2cos\frac{1}{2}(k_{1}-k_{2})cos\frac{1}{2}(k_{1}+k_{2})\nonumber \\
 & = & 2-2cos\frac{1}{2}(k_{1}-k_{2})cos\frac{k}{2}\label{123} 
\end{eqnarray}
The value of k lies in the range \( -\pi \leq k\leq \pi  \) . For each value
of k, \( k_{1}-k_{2} \) lies in the range \( 0\leq (k_{1}-k_{2})\leq 2\pi  \)
. Thus if one plots \( \epsilon _{I} \) against k, one gets a continuum of
scattering states (Figure 7).

Now, the minimum value of \( \epsilon _{I} \) is

\begin{equation}
\label{124}
\epsilon ^{min}_{I}=2-2cos\frac{k}{2}
\end{equation}
 From (122),

\begin{equation}
\label{125}
\epsilon _{II}=1-cos^{2}\frac{k}{2}
\end{equation}
 Thus,

\begin{equation}
\label{126}
\frac{\epsilon _{II}}{\epsilon ^{min}_{I}}=\frac{1}{2}(1+cos\frac{k}{2})\leq 1
\end{equation}
 We find that \( \epsilon _{II<}\epsilon _{I}^{min} \) for k \( \neq  \)0.
For k= 0, the bound state is degenerate with an eigenstate belonging to the
continuum of scattering states and having the energy eigenvalue zero. Thus there
are N-1 bound state solutions. 

We now consider the wave functions in Cases I and II. Fos Case I, \( k_{1},k_{2} \)
and \( \phi  \) are real and

\begin{equation}
\label{127}
a(m_{1},m_{2})=e^{i(k_{1}m_{1}+k_{2}m_{2}+\phi /2)}+e^{i(k_{2}m_{1}+k_{1}m_{2}-\phi /2)}
\end{equation}
The state describes the scattering of spin waves. The `phase shift' \( \phi  \)
results from the mutual interaction between the spin waves. Since spin operators
at two different sites commute, the excitations behave like bosons and the amplitude
\( a(m_{1},m_{2}) \) is properly symmetrized to reflect the bosonic nature
of the spin waves. In Case II, from (120) and with \( \psi  \) = 0 and \( \chi  \)=Nv,
we get

\begin{eqnarray}
a(m_{1},m_{2}) & = & e^{i\phi /2}e^{i(k_{1}m_{1}+k_{2}m_{2})}+e^{-i\phi /2}e^{i(k_{2}m_{1}+k_{1}m_{2})}\nonumber \\
 & = & e^{ik/2(m_{1}+m_{2})}\left[ e^{-Nv/2-v(m_{1}-m_{2})}+e^{Nv/2+v(m_{1}-m_{2})}\right] \nonumber \\
 & = & 2e^{i\frac{k}{2}(m_{1}+m_{2})}coshv\left[ \frac{N}{2}-(m_{2}-m_{1})\right] \label{128} 
\end{eqnarray}
If we normalise these states, we see that the two reversed spins tend to be
localised at n.n. positions; \( \mid a(m_{1},m_{2})\mid  \)is a maximum for
\( m_{2}=m_{1}+1 \) and has an exponential decay for \( m_{2}>m_{1}+1 \) .
The width of the bound state is given by v and u is the velocity of the centre
of mass of the bound complex in the chain. The Case II solutions represent the
bound states of two spin waves. 

Finally, we consider Case III,i.e., \( \lambda _{1}=(\lambda -1)/2 \) and \( \lambda _{2}=(\lambda +1)/2 \)
(\( \lambda  \) odd). From (116), we get

\begin{eqnarray}
Nk_{1}-\phi  & = & \pi (\lambda -1)\nonumber \\
Nk_{2}+\phi  & = & \pi (\lambda +1)\label{129} 
\end{eqnarray}
A solution of (129) is \( k_{1}=k_{2} \) and \( \phi =\pi  \). Then from (127),

\begin{equation}
\label{130}
a(m_{1},m_{2})=e^{ik/2(m_{1}+m_{2})}\left[ exp(i\pi /2)+exp(-i\pi /2\right] =0
\end{equation}
For complex \( k_{1},k_{2} \) and \( \phi  \), \( \psi =\pi  \) and \( \chi  \)
= Nv from (120). For N \( \rightarrow \infty  \) , we obtain the same solutions
as we find in Case II. 

For the general case n = r, Bethe proposed a form for the amplitude \( a(\{m\}) \)
(the celebrated BA) which is a generalisation of the amplitude in the n =2 case.

\begin{equation}
\label{131}
a(m_{1},m_{2},......,m_{r})=\sum _{P}exp\left[ i\sum ^{r}_{j=1}k_{Pj}m_{j}+\frac{1}{2}i\sum ^{1,r}_{j<l}\phi _{Pj,Pl}\right] 
\end{equation}
The sum over P denotes a sum over all the r! permutations of the integers 1,2,...,r,
Pj is the image of j under the permutation P. Each term in (131) has r plane
waves scattering against one another. For each pair of plane waves, there is
a phase shift \( \phi _{j,l} \) . The sum over permutations is in accordance
with the bosonic nature of the waves, the spin waves, propagating along the
chain. For r spin deviations several possibilities occur: none of the spin deviations
occur at n.n. sites, only two of the spin deviations occur at n.n. sites, more
than a single pair of spin deviations occur at n.n. sites, three spin deviations
occur at n.n. sites and so on. However, only the first two possibilities need
to be considered to solve the eigenvalue problem, the other possibilities do
not give any new information {[}51{]}. For r= 3, let us now write down the amplitude
\( a(m_{1},m_{2},m_{3}) \) in the BA form given by (131).Since r =3, the sum
contains 3! = 6 terms.

\begin{eqnarray}
a(m_{1},m_{2},m_{3}) & = & e^{i(k_{1}m_{1}+k_{2}m_{2}+k_{3}m_{3})+i/2(\phi _{12}+\phi _{13}+\phi _{23})}\nonumber \\
 & + & e^{i(k_{1}m_{1}+k_{3}m_{2}+k_{2}m_{3})+i/2(\phi _{12}+\phi _{13}+\phi _{32})}\nonumber \\
 & + & e^{i(k_{2}m_{1}+k_{1}m_{2}+k_{3}m_{3})+i/2(\phi _{21}+\phi _{13}+\phi _{23})}\nonumber \\
 & + & e^{i(k_{3}m_{1}+k_{1}m_{2}+k_{2}m_{3})+i/2(\phi _{12}+\phi _{31}+\phi _{32})}\nonumber \\
 & + & e^{i(k_{2}m_{1}+k_{3}m_{2}+k_{1}m_{3})+i/2(\phi _{21}+\phi _{31}+\phi _{23})}\nonumber \\
 & + & e^{i(k_{3}m_{1}+k_{2}m_{2}+k_{1}m_{3})+i/2(\phi _{21}+\phi _{31}+\phi _{32})}\label{132} 
\end{eqnarray}
For the general case n = r, i.e., r up-spins, one can proceed as in the n =
2 case and obtain the following equations:

\begin{equation}
\label{133}
\epsilon =\sum ^{r}_{i=1}(1-cosk_{i})
\end{equation}

\begin{equation}
\label{134}
2cot\frac{1}{2}\phi _{j.l}=cot\frac{k_{j}}{2}-cot\frac{k_{l}}{2},-\pi \leq \phi _{j.l}\leq \pi 
\end{equation}

\begin{equation}
\label{135}
Nk_{i}=2\pi \lambda _{i}+\sum _{j}\phi _{ij}
\end{equation}
The Eqs. (134) are r(r-1)/2 in number. Since \( \phi _{jl}=-\phi _{lj} \) ,
there are only r(r-1)/2 distinct \( \phi  \)'s. The Eqs. (135) are r in number.
These r equations along with the r(r-1)/2 Eqs. of the type (134) give r(r+1)/2
equations in as many unknowns. Thus the equations can be expected to have solutions.
Bethe showed that they also give the correct number of solutions. 

We now consider a particular type of solution, in which the r deviated spins
form a bound state and move together. The following simplified analysis has
been given by Ovchinnikov {[}52{]}. One assumes that r <\textcompwordmark{}<N
and that only the phases \( \phi _{12},\phi _{23},.....,\phi _{r-1,r} \) are
large. One may put \( Im\phi _{l-1,l}>0 \) without any loss of generality.
Then from (135),

\begin{eqnarray}
Im\phi _{12} & = & NImk_{1}=Nx_{1}\nonumber \\
Im(\phi _{23}-\phi _{12}) & = & NImk_{2}=Nx_{2}\nonumber \\
Im\phi _{r-1,r} & = & Nx_{r}\label{136} 
\end{eqnarray}

One has 

\( \sum ^{r}_{i=1}x_{i}\geq 0 \) for all r. Substituting in (Eq. (134)), one
gets to an accuracy \( e^{-N} \)

\begin{equation}
\label{137}
2cot\frac{\phi _{12}}{2}=2i\frac{e^{i(Re\phi _{12}+iIm\phi _{12})}+e^{-i(Re\phi _{12}+iIm\phi _{12})}}{e^{i(Re\phi _{12}+iIm\phi _{12})}-e^{-i(Re\phi _{12}+iIm\phi _{12})}}=-2i
\end{equation}
 Thus,

\begin{eqnarray}
-2i & = & cot\frac{k_{1}}{2}-cot\frac{k_{2}}{2}\nonumber \\
-2i & = & cot\frac{k_{2}}{2}-cot\frac{k_{3}}{2}etc.\label{138} 
\end{eqnarray}
So, 

\begin{equation}
\label{139}
cot\frac{k_{l}}{2}=2i+cot\frac{k_{l-1}}{2}
\end{equation}
The solution is given by

\begin{equation}
\label{140}
cot\frac{k_{l}}{2}=2il+C
\end{equation}
To determine C, introduce the total momentum of the deviated up-spins

\[
k=\sum ^{r}_{i=1}k_{i}=\frac{2\pi }{N}\sum ^{r}_{i=1}\lambda _{i}\]
The wave function \( \psi  \) is multiplied by \( e^{ik} \) due to a shift
\( m_{i}\rightarrow m_{i}+1 \) and the energy levels are characterised by k.
One gets

\begin{eqnarray}
C & = & \frac{2ri-i(e^{ik}-1)}{e^{ik}-1}\nonumber \\
e^{ikl} & = & \frac{r+l(e^{ik}-1)}{r+(l-1)(e^{ik}-1)}\label{141} \\
\epsilon  & = & \sum ^{r}_{l=1}(1-cosk_{l})\nonumber \\
 & = & \frac{1}{r}(1-cosk)\nonumber 
\end{eqnarray}
The results can be generalised to the Hamiltonian with longitudinal anisotropy

\begin{equation}
\label{142}
H(J,\sigma )=-J\sum ^{N}_{i=1}\left[ S^{z}_{i}S^{z}_{i+1}+\sigma (S^{x}_{i}S^{x}_{i+1}+S^{y}_{i}S^{y}_{i+1})\right] 
\end{equation}
The anisotropy parameter \( \sigma  \) lies between 0 and 1. The multimagnon
bound states were first detected in the quasi-one-dimensional magnetic system
\( CoCl_{2}.2H_{2}O \) at pumped helium temperature in high magnetic fields
by far infrared spectroscopy {[}53{]}. Later improvements {[}54{]} made use
of infrared HCN/DCN lasers, the high intensity of which made possible observation
of even 14 magnon bound states.

We now describe the calculation of the ground state energy of the HAFM Hamiltonian
using the BA. The sign of the exchange integral changes from - J to J (J > 0)
and \( \epsilon =\frac{(E-JN/4)}{J} \) . The BA Eqns. are still given by (133)-(135),
only there is an overall `minus' sign on the r.h.s. of (133). In the AFM ground
state, N/2 spins are up (r=N/2) and N/2 spins down. The numbers \( \lambda _{i} \)
's are ordered as 

\begin{equation}
\label{143}
0<\lambda _{1}\leq \lambda _{2}\leq \lambda _{3}\leq .......
\end{equation}
 Again, \( \lambda _{j+1}\geq \lambda _{j}+2 \) , for real \( k_{i} \) 's
(Case I ). There is a unique choice of the \( \lambda _{i} \) 's as

\begin{equation}
\label{144}
\lambda _{1}=1,\lambda _{2}=3,\lambda _{3}=5,....,\lambda _{\frac{N}{2}}=N-1
\end{equation}
The ground state has total spin S = 0 and is non-degenerate. The total ground
state wave vector is

\begin{equation}
\label{145}
P=\sum ^{N/2}_{i=1}k_{i}=\frac{2\pi }{N}(1+3+........+(N-1))=\frac{\pi }{2}N
\end{equation}
If N = 4m, m positive, the wave vector is 0 (mod \( 2\pi  \) ); if N = 4m+2,
it is \( \pi  \) (mod 2\( \pi  \)). The ground state energy is calculated
by a passage to the continuum limit:

\begin{equation}
\label{146}
\lambda _{j}=\frac{2j-1}{N}\rightarrow x
\end{equation}
\begin{equation}
\label{147}
\frac{1}{N}\sum ^{N/2}_{l=1}\phi _{jl}\rightarrow \frac{1}{2}\int ^{1}_{0}\phi (x,y)dy
\end{equation}
 Eqs. (133)-(135) become 

\begin{equation}
\label{148}
2cot\frac{1}{2}\phi (x,y)=cot\frac{1}{2}k(x)-cot\frac{1}{2}k(y)
\end{equation}
 
\begin{equation}
\label{149}
k(x)=2\pi x+\frac{1}{2}\int ^{1}_{0}\phi (x,y)dy
\end{equation}
 
\begin{eqnarray}
\epsilon  & = & -\sum ^{N/2}_{j=1}(1-cosk_{j})\nonumber \\
 & = & -\frac{N}{2}\int ^{1}_{0}(1-cosk(x))dx\label{150} 
\end{eqnarray}
 We note that when x = y, \( \phi  \) jumps from \( \pi  \) to - \( \pi  \)
and

\begin{equation}
\label{151}
\frac{\partial \phi (x,y)}{\partial x}=-2\pi \delta (x-y)+2\frac{d}{dx}(cot^{-1}\frac{B}{A})
\end{equation}
 where A =2 and \( B=cot\frac{k(x)}{2}-cot\frac{k(y)}{2} \) . Thus from (149),
we get 

\begin{equation}
\label{152}
\frac{dk}{dx}=\pi +cosec^{2}(\frac{k}{2})\frac{dk}{dx}\int ^{1}_{0}\frac{dy}{4+\{cot\frac{1}{2}k(x)-cot\frac{1}{2}k(y)\}^{2}}
\end{equation}
Put

\begin{equation}
\label{153}
g=cot\frac{1}{2}k
\end{equation}
and

\begin{equation}
\label{154}
\rho _{0}(g)=-\frac{dx}{dy}
\end{equation}
 Then from (152), we get the integral equation derived by Hulthén

\begin{equation}
\label{155}
\frac{2}{\pi (1+g^{2})}=\rho _{0}(g)+\frac{2}{\pi }\int ^{+\infty }_{-\infty }\frac{\rho _{0}(g^{\prime })dg^{\prime }}{(4+(g-g^{\prime })^{2})}
\end{equation}
 The integral equation can be solved by the method of iterations,

\begin{equation}
\label{156}
\rho _{0}(g)=\frac{2}{\pi (1+g^{2})}-\lambda \int ^{+\infty }_{-\infty }\frac{\rho _{0}(g^{\prime })dg^{\prime }}{(4+(g-g^{\prime })^{2})}
\end{equation}
(at the end of the calculation \( \lambda  \) is put equal to 2/\( \pi  \)).
On iteration, (156) becomes

\begin{eqnarray}
\rho _{0}(g) & = & \frac{2}{\pi (1+g^{2})}-\lambda \int ^{+\infty }_{-\infty }\frac{dg_{1}}{(g_{1}-g)^{2}+4}\nonumber \\
 & \times  & \left[ \frac{2}{\pi (1+g_{1}^{2})}-\lambda \int ^{+\infty }_{-\infty }\frac{\rho _{0}(g_{2})dg_{2}}{(4+(g_{2}-g_{1})^{2})}\right] \label{157} 
\end{eqnarray}
 One can continue to iterate (157) to generate higher order terms in \( \lambda  \).

\begin{eqnarray}
\rho _{0}(g) & = & \frac{2}{\pi (1+g^{2})}-\lambda \frac{2}{\pi }\int ^{+\infty }_{-\infty }\frac{dg_{1}}{(4+(g_{1}-g)^{2})(g^{2}_{1}+1)}\nonumber \\
 & + & \lambda ^{2}\frac{2}{\pi }\int ^{+\infty }_{-\infty }\frac{dg_{1}dg_{2}}{(4+(g_{1}-g)^{2})(4+(g_{2}-g_{1})^{2})(g^{2}_{2}+1)}\nonumber \\
 & - & \lambda ^{3}\int ...................\label{158} 
\end{eqnarray}
One can calculate the integrals in (158) by the method of residues. For example,

\begin{equation}
\label{159}
\frac{2}{\pi }\int ^{+\infty }_{-\infty }\frac{dg_{1}}{(4+(g_{1}-g)^{2})(g^{2}_{1}+1)}=\frac{3}{g^{2}+9}
\end{equation}
 Putting \( \lambda =\frac{2}{\pi } \), we finally obtain the solution of the
integral equation (156) as

\begin{equation}
\label{160}
\rho _{0}(g)=\frac{2}{\pi }\left[ \frac{1}{g^{2}+1}-\frac{3}{g^{2}+3^{2}}+\frac{5}{g^{2}+5^{2}}-..........\right] 
\end{equation}
 The series in (160) is uniformly convergent and one gets,

\begin{equation}
\label{161}
\rho _{0}(g)=\frac{1}{2}sech\frac{\pi g}{2}
\end{equation}
 Thus,

\begin{equation}
\label{162}
\epsilon =-N\int ^{+\infty }_{-\infty }\frac{\rho _{0}(g)}{1+g^{2}}dg=-Nln2
\end{equation}
 In absolute measure, the ground state energy \( E_{g} \) is

\begin{equation}
\label{163}
E_{g}=\frac{NJ}{4}-JNln2
\end{equation}
 The low-lying excitation spectrum has been calculated by des Cloizeaux and
Pearson (dCP) {[}55{]} by making appropriate changes in the distribution of
\( \lambda _{i}'s \) in the ground state. The spectrum is given by

\begin{equation}
\label{164}
\epsilon =\frac{\pi }{2}\mid sink\mid ,-\pi \leq k\leq \pi 
\end{equation}
 for spin 1 states. The wave vector k is measured w.r.t. that of the ground
state. A more rigorous calculation of the low-lying excitation spectrum has
been given by Faddeev and Takhtajan {[}56{]}. There are S =1 as well as S =
0 states. We give a qualitative description of the excitation spectrum, for
details Ref. {[}56{]} should be consulted. The energy of the low-lying excited
states can be written as \( E(k_{1},k_{2})=\epsilon (k_{1})+\epsilon (k_{2}) \)
with \( \epsilon (k_{i})=\frac{\pi }{2}sink_{i} \) and total momentum k = \( k_{1}+k_{2} \)
. At a fixed total momentum k, one gets a continuum of scattering states (Figure
8). The lower boundary of the continuum is given by the dCP spectrum (one of
the \( k_{i}' \)s = 0). The upper boundary is obtained for \( k_{1}=k_{2}=k/2 \)
and

\begin{equation}
\label{165}
\epsilon ^{U}_{k}=\pi \mid sin\frac{k}{2}\mid 
\end{equation}
The energy-momentum relations suggest that the low-lying spectrum is actually a combination of two elementary excitations known as spinons. The energies and the momenta of the spinons just add up, showing that they do not interact. A spinon is a S = 1/2 object, so on combination they give rise to both S=1 and S=0 states. In the Heisenberg model, the spinons are only noninteracting in the thermodynamic limit N\( \rightarrow \infty  \). For an even number N of sites, the total spin is always an integer, so that the spins are always excited in pairs. The spinons can be visualised as kinks in the AFM order parameter (Figure 9). Due to the exchange interaction, the individual spinons get delocalized into plane wave states. Inelastic neutron scattering study of the linear chain S=1/2 HAFM compound \( KCuF_{3} \) has confirmed the existence of unbound spinon pair excitations [57]. In the case of integer spin chains, the spinons are bound and the excitation spectrum consists of spin-w!
ave-like modes exhibiting the Haldane gap. The BA technique described in this Section is the one originally proposed by Bethe. There is an algebraic version of the BA which is in wide use and which gives the same final results as the earlier technique. For an introduction to the algebraic BA method, see the Refs. [58,59].

\newpage

\bf{Figure Captions}

Figure 1. Ising spins on a triangular plaquette.

Figure 2. An example of a RVB state.

Figure 3. The Shastry-Sutherland model.

Figure 4. Five types of interactions in the \( J_{1}-J_{2}-J_{3}-J_{4}-J_{5} \)
model.

Figure 5. Four columnar dimer states.

Figure 6. A two-chain spin ladder.

Figure 7. The continuum of scattering states (solid lines) and the bound state 

(dotted line) of two magnons.

Figure 8. Continuum of scattering states for the S = 1/2 Heisenberg AFM chain.

Figure 9. A two-spinon configuration in an AFM chain.

\end{document}